\documentstyle[12pt,epsfig]{article}
\begin{document}
\setlength{\baselineskip}{0.30in}
\newcommand{\nc}{\newcommand}
\newcommand{\beq}{\begin{equation}}
\newcommand{\eeq}{\end{equation}}
\newcommand{\be}{\begin{eqnarray}}
\newcommand{\ee}{\end{eqnarray}}
\newcommand{\num}{\nu_\mu}
\newcommand{\nue}{\nu_e}
\newcommand{\nut}{\nu_\tau}
\newcommand{\nnt}{n_{\nu_\tau}}
\newcommand{\rnt}{\rho_{\nu_\tau}}
\newcommand{\mnt}{m_{\nu_\tau}}
\newcommand{\tnt}{\tau_{\nu_\tau}}
\newcommand{\bi}{\bibitem}
\newcommand{\rar}{\rightarrow}
\newcommand{\lar}{\leftarrow}
\newcommand{\lrar}{\leftrightarrow}
\newcommand{\dm}{\delta m^2}
\newcommand{\so}{\, \mbox{sin}\Omega}
\newcommand{\co}{\, \mbox{cos}\Omega}
\newcommand{\sotil}{\, \mbox{sin}\tilde\Omega}
\newcommand{\cotil}{\, \mbox{cos}\tilde\Omega}

\begin{small}
{\hbox to\hsize{ \hfill TAC-1999-018}}
{\hbox to\hsize{ \hfill SISSA 124/99/EP}}
\end{small}
\begin{center}
\vglue .06in
{\Large \bf {Neutrino oscillations in the early universe: \\
How large lepton asymmetry can be generated?}}
\bigskip
\\{\bf A.D. Dolgov
\footnote{Also: ITEP, Bol. Cheremushkinskaya 25, Moscow 113259, Russia.}
\footnote{e-mail: {\tt dolgov@tac.dk}},
S.H. Hansen\footnote{e-mail: {\tt sthansen@tac.dk}}
 \\[.05in]
{\it{Teoretisk Astrofysik Center\\
 Juliane Maries Vej 30, DK-2100, Copenhagen, Denmark}
}}
\\{\bf S. Pastor \footnote{e-mail: {\tt pastor@sissa.it}}} \\
{\it{SISSA--ISAS and INFN, Sezione di Trieste\\
Via Beirut 2-4, I-34013 Trieste, Italy}
}
\\{\bf D.V. Semikoz \footnote{e-mail: {\tt semikoz@ms2.inr.ac.ru}}} \\
{\it{Max-Planck-Institut f\"ur Physik (Werner-Heisenberg-Institut)\\
F\"ohringer Ring 6, 80805 M\"unchen, Germany\\
and\\
Institute of Nuclear Research of the Russian Academy of Sciences\\
60th October Anniversary Prospect 7a, Moscow 117312, Russia}}
\\[.40in]
\end{center}

\begin{abstract}
The lepton asymmetry that could be generated in the early universe through
oscillations of active to sterile neutrinos is calculated (almost)
analytically for small mixing angles, sin$2 \theta <10^{-2}$. It is shown
that for a mass squared difference, $\delta m^2=-1$ eV$^2$ it may rise at
most by 6 orders of magnitude from the initial ``normal'' value
$\sim 10^{-10}$, since the back-reaction from the refraction index terminates
this rise while the asymmetry is still small. Only for very large mass
differences, $|\delta m^2|\sim 10^9$ eV$^2$, the lepton asymmetry could reach
a significant magnitude exceeding 0.1. 
\end{abstract}

Neutrino oscillations in the early universe differ from e.g. solar
oscillations in two important aspects. First, one cannot neglect
neutrino annihilation or scattering in the medium for a rather large
and physically interesting range of parameters (mixing angle, $\sin 2
\theta$, and mass squared difference, $\delta m^2$). These processes
break the coherence of neutrino propagation and that is why
considerations in terms of wave functions become impossible and one
has to use the density matrix formalism~\cite{dolgov81,stodolsky87}.
Kinetic equations for the density matrix of oscillating neutrinos with
the account of the so called second order effects (proportional to the
second power of the Fermi coupling constant, $G_F$) were derived in
refs.~\cite{dolgov81,raffelt93, sigl93}. Second, neutrino oscillations
in the primeval plasma could modify the plasma properties, in
particular the refraction index, and this in turn would influence the
oscillations, so that the problem becomes highly nonlinear.  The
refraction index of oscillating neutrinos in the cosmic plasma was
calculated in ref.~\cite{notzold88}.  An important feature of the
refraction index is that it contains terms proportional to the charge
asymmetry of the cosmic plasma. The effective potential of a standard
neutrino of flavor $a$ can be written as
\be
V_{eff}^a =
\pm C_1 \eta G_FT^3 + C_2^a \frac{G^2_F T^4 E}{\alpha} ~,
\label{nref}
\ee
where $E$ is the neutrino energy, $T$ is the temperature of the
plasma, $G_F=1.166\cdot 10^{-5}$ GeV$^{-2}$ is the Fermi coupling
constant, $\alpha=1/137$ is the fine structure constant, and the signs
``$\pm$'' refer to anti-neutrinos and neutrinos respectively (this
choice of sign describes the helicity state, negative for $\nu$ and
positive for $\bar\nu$). According to ref.~\cite{notzold88} the
coefficients $C_j$ are: $C_1 \approx 0.95$, $C_2^e \approx 0.61$ and
$C_2^{\mu,\tau} \approx 0.17$.  These values are true in the case of
thermal equilibrium, and otherwise these coefficients are some
integrals over the distribution functions.  In ref.~\cite{notzold88}
the coefficient $C_1$ was calculated using the present value of
the asymmetry $\eta$, which differs from its value in the early Universe
(before $e^+e^-$ annihilation increased the photon temperature) by
a factor $11/4$. In our calculations we took $C_1=0.345$.  The charge
asymmetry, $\eta$, is defined as the ratio of the difference between
particle-antiparticle number densities to the number density of
photons.  The individual contributions to $\eta$ from different
particle species are the following:
\be
\eta &=& 2\eta_{\nue} +\eta_{\num} + \eta_{\nut} +\eta_{e}-\eta_{n}/2 \,\,\,
 ( {\rm for} \,\, \nue)~,
\label{etanue} \\
\eta &=& 2\eta_{\num} +\eta_{\nue} + \eta_{\nut} - \eta_{n}/2\,\,\,
({\rm for} \,\, \num)~,
\label{etanumu}
\ee
and $\eta$ for $\nut$ is obtained from eq.~(\ref{etanumu}) by the interchange
$\mu \lrar \tau$.

The different magnitude of the refraction indices for neutrinos and
anti-neutrinos may result in more favorable conditions for $\nu_a \rar
\nu_s$ oscillations compared with $\bar \nu_a \rar \bar \nu_s$
oscillation, where $\nu_a$ is an active neutrino, $a = e,\mu, \tau$,
and $\nu_s$ is a sterile one.  Since more $\nu_a$ than $\bar\nu_a$
would be transformed into sterile ones, the lepton asymmetry in the
sector of active neutrinos would rise and this would further amplify
the process. The possibility of such instability was noticed in
ref.~\cite{barbieri91}, but there it was found, on the basis of
simplified considerations, that the rise is stabilized when non-linear
terms in the refraction index become non-negligible, and thus it was
concluded, that ``no large chemical potential will be generated in any
point of the parameter space''. A similar statement of a small
asymmetry was made in ref.~\cite{enkvist90}.
This conclusion was reconsidered in ref.~\cite{foot96} (see also
refs.~\cite{foot97,footAP99}) where it was argued that a very large
asymmetry, even close to 1, may be generated by the oscillations. An
even more striking statement was made in ref.~\cite{shi96}, that the
asymmetry may not only be large, but may have a chaotic sign, so that
even domains with different signs of the asymmetry may be
formed~\cite{shi99}. Similar results were obtained in
the recent paper~\cite{enqvist99},
namely, that the asymmetry may reach large values and in some ranges of
parameters its sign may be chaotic, while in ref.~\cite{dibari99}
chaoticity was not observed. However, these results were derived in a
simplified way after averaging some essential quantities over neutrino
momenta or through a solution of momentum dependent but simplified
equations. At the same time, in ``brute force'' numerical calculations
applied to this problem, it is very difficult to distinguish between
the real effect and a computational instability. Technically there is an
essential difference between exact momentum dependent calculations and
the momentum averaged ones. In the last case one has to solve a set of
ordinary differential equations, while in the former case the
corresponding equations are integro-differential ones (see
e.g. ref.~\cite{dolgov97}, where exact kinetic equations were
accurately solved for non-oscillating neutrinos).  Momentum dependent exact
numerical calculations were done in a series of papers~\cite{kirilova97}.
It was shown there that the asymmetry may rise by several (3--4) orders
of magnitude but still remains small in the range of
parameters $|\dm| < 10^{-7}$ eV$^2$, assuming a small initial
asymmetry, $10^{-10}$.  Outside this range of parameters the
calculations became unstable and no definite result was obtained. We
extended the domain of stable computation to a somewhat larger range
of parameters~\cite{dolgov99}:
\be
|\dm| < 10^{-6} ~{\rm eV}^2~,
\label{dmtheta}
\ee
and have also not found any large generation of charge asymmetry. In
the coinciding range of parameters our results are in a reasonable
agreement with those of ref.~\cite{kirilova97}.  The attempts to
extend the range (\ref{dmtheta}) maintaining the stability of the
computational process demanded a huge increase of computer time,
because otherwise the results were chaotic in sign and showed a quickly
rising asymmetry.  Thus it seems that further attempts to extend stable
numerical calculations to a wider parameter range, in which the huge
amplification of the asymmetry was found, will be fruitless and one
should try to transform the kinetic equations for the oscillating
neutrinos analytically to such a form that will permit a numerical
solution. In what follows we have achieved this goal and reduced the
problem to the solution of an (almost) ordinary differential equation
for the neutrino charge asymmetry, which is easy to solve
numerically.

Before presenting the actual calculations we will briefly describe our
procedure so that it will be easier to follow it.  We start from the
usual equations for the evolution of the neutrino (as well as
anti-neutrino) density matrix in a cosmological environment. We twice
introduce new variables, first $x$ and $y$ given by eq.~(\ref{xy}) so
that the evolution operator in the l.h.s.~of the kinetic equations
would depend only on one variable $x$, and second, we rewrite all the
equations in terms of $\tau$ given by eq.~(\ref{eta}).  It is a
natural variable for the description of the behavior of the density
matrix near the MSW resonance, where it takes the value $\tau=1$, in
the limit of a vanishing contribution of charge asymmetry to the
refraction index.  As we see in what follows almost all coefficient
functions in the kinetic equations depend only on $\tau$ except for
the charge asymmetric potential $V$ in eq.~(\ref{newv}) that depends
on two variables $\tau$ and $y$.  Using the variable $\tau$ permits to
factor out the large parameter $Q$ in eq.~(\ref{Q}) related to a large
frequency of oscillations. This is true for neutrino mass differences
above $10^{-7}$ eV$^2$. Quick oscillations make numerical computations
very difficult. Fortunately one may analytically separate the quickly
varying functions and make an expansion in terms of $1/Q$.  An
essential technical step is to consider charge symmetric and
antisymmetric elements of the density matrix, $\rho \pm \bar
\rho$. Antisymmetric combination directly enters the refraction index (see
eqs.~(\ref{nref},\ref{VandZ})) and working with symmetric and
antisymmetric functions permits to derive a closed equation for the
evolution of the asymmetry. As a first step we formally solve
equations (\ref{firstasyms})--(\ref{firstasyml}) for the antisymmetric
elements of the density matrix in terms of unknown symmetric functions and
the integrated charge asymmetry $Z$ in eq.~(\ref{VandZ}). In the limit
of large $Q$ the corresponding differential equations allow a simple
algebraic solution. As a second step we substitute the obtained
expressions into the charge symmetric equations. For the latter we
find eigenfunctions that are formal solutions of these equations in
the case of constant coefficients. However since the latter are not
constant we obtain a system of differential equations for the
coefficient functions in the expansion of the solution in terms of the
eigenvectors. The equations for these coefficient functions are quite
simple and can be solved numerically and analytically (both approaches
give very close results). After that we substitute the found
solutions, which contain an unknown charge asymmetry $Z$, back into
the antisymmetric equations and after integration over momentum of
both sides of the equation for $(\rho_{aa}'-\bar\rho_{aa}')$ we obtain
a closed differential equation for the charge asymmetry $Z$. The
latter is a function of a single variable $q=y\tau$ and it can be
relatively easily solved numerically.  Now we will describe the same
in more detail.

The basic equations governing evolution of density matrix are:
\be
i(\partial_t -Hp\partial_p) \rho_{aa}
&=& F_0(\rho_{sa}-\rho_{as})/2 -i \Gamma_0 (\rho_{aa}-f_{eq})~,
\label{dotrhoaa} \\
i(\partial_t -Hp\partial_p)  \rho_{ss}
&=& -F_0(\rho_{sa}-\rho_{as})/2~,
\label{dotrhoss} \\
i(\partial_t -Hp\partial_p) \rho_{as} &=&
W_0\rho_{as} +F_0(\rho_{ss}-\rho_{aa})/2-
i\Gamma_1 \rho_{as} ~,
\label{dotrhoas}\\
i(\partial_t -Hp\partial_p) \rho_{sa} &=& -W_0\rho_{sa} -
F_0(\rho_{ss}-\rho_{aa})/2- i\Gamma_1 \rho_{sa}  ~,
\label{dotrhosa}
\ee
where $a$ and $s$ mean ``active'' and ``sterile'' respectively, $F_0=\dm
\sin 2\theta / 2E$, $W_0= \dm\cos 2\theta /2E + V_{eff}^a$,
$H=\sqrt{8\pi \rho_{tot}/3M_p^2}$ is the Hubble parameter, $p$ is the
neutrino momentum, and $f_{eq}$ is the equilibrium Fermi distribution
function:
\be
f_{eq}= \left[ \exp\left(E/T\right) +1\right]^{-1}~.
\label{feq}
\ee
More precisely instead of the equilibrium function $f_{eq}$ one should
use the one with a non-zero chemical potential, because scattering and
annihilation processes do not change lepton number. However if the
asymmetry is not large the difference between them is not important
for our calculations.

The anti-neutrino density matrix satisfies the similar set of equations
with the opposite sign of the antisymmetric term in $V_{eff}^a$ and with
a slight difference in damping factors that is proportional to the
lepton asymmetry.

Equations (\ref{dotrhoaa}-\ref{dotrhosa})
account exactly for the first order terms described by
the refraction index, while the second order terms describing coherence
breaking are approximately modeled by the damping coefficients $\Gamma_j$.
The latter are equal to~\cite{harris}:
\be
\Gamma_0 = 2\Gamma_1  = g_a \frac{180 \zeta(3)}{7 \pi ^4}
\, G_F^2 T^4 p  ~.
\label{gammaj1}
\ee
In general the coefficient $g_a(p)$ is a momentum-dependent
function, but in the approximation of neglecting $[1-f]$ factors in the
electro-weak collision rates it becomes a constant~\cite{bell99} that
corresponds to $g_{\nu_e} \simeq 4$ and $g_{\nu_\mu,\mu_\tau} \simeq
2.9$ \cite{enqvist92}.  In the following we will use a more accurate value
of $g_a$, which comes from the thermal average of the complete
electro-weak rates (with factors $[1-f]$ included), which we
calculated numerically from our Standard Model code
\cite{dolgov97}. This gives us $g_{\nu_e} \simeq 3.56$
and $g_{\nu_\mu,\mu_\tau} \simeq 2.5$.
The indices sub-0 are here prescribed to the coefficient functions to
distinguish them from the similar ones after dividing by $Hx$ (see
below).

It is convenient to introduce new variables:
\be
x=m_0 R(t)\,\, {\rm and}\,\,   y=p R(t)~,
\label{xy}
\ee
where $R(t)$ is the cosmological scale factor so that $H=\dot
R/R$ and $m_0$ is an arbitrary mass (just normalization), we choose
$m_0 =1$ MeV. In the approximation that we will work, we assume that
$\dot T =-HT$, so that we can take $R=1/T$. In terms of these
variables the differential operator $(\partial_t -Hp\partial_p)$
transforms to $Hx\partial_x$. We will normalize the density matrix
elements\footnote{Other authors find it convenient to express this density matrix
formalism in terms of Pauli matrices and a polarization vector, $\vec P =
(P_x, P_y, P_z)$, such that:
\[
\rho \equiv \frac{P_0}{2} \left[ 1 + \vec P \cdot \vec \sigma \right]~,
\]
in such a way that $P_0P_z = 1$ means that all the neutrinos are
$\nu_e$, and we have $P_0P_x = f_{eq}h$, $P_0P_y = -f _{eq} l$,
$P_0P_z = f_{eq} (a-s)$ and $P_0 = f_{eq} (2 + a+s)$.} to the
equilibrium distribution:
\be
\rho_{aa} &=& f_{eq}(y) [1+a(x,y)],\,\, \rho_{ss} = f_{eq}(y) [1+s(x,y)]~, \\
\label{rhoaa}
\rho_{as} &=& \rho_{sa}^* = f_{eq}(y)[h(x,y)+i l(x,y)]~,
\label{hil}
\ee
and the neutrino mass difference $\delta m^2$ to eV$^2$.

As the next step we will take the sum and difference of
eqs.~(\ref{dotrhoaa})-(\ref{dotrhosa}) for $\nu$ and $\bar\nu$. The
corresponding equations have the following form:
\be
s_{\pm}' &=& F l_{\pm}~, \label{firstasyms} \\
a_{\pm}' &=& - F l_{\pm} - 2 \gamma_+ a_{\pm} - 2 \gamma_- a_{\mp} ~,\\
h_{\pm}' &=& U l_{\pm} - V Z l_{\mp} - \gamma_+ h_{\pm} -
\gamma_- h_{\mp} ~,\\
l_{\pm}' &=&\frac{F}{2}(a_{\pm} - s_{\pm}) - U h_{\pm} + V Z h_{\mp} -
\gamma _+ l _{\pm}
- \gamma_- l_{\mp}~,
\label{firstasyml}
\ee
where $a_{\pm} = (a \pm {\bar a})/2$ etc, and the prime means
differentiation with respect to $x$.  We have used $W=U \pm VZ$,
$\gamma =\Gamma_1/Hx$, and $\gamma_{\pm} =(\gamma \pm \bar\gamma)/2$,
where $\gamma_-$ parameterizes the difference of interaction rates
between neutrino and anti-neutrinos, which is proportional to the
neutrino asymmetry. With the approximation\footnote{
This approximation is valid for high temperatures $T > 1$ MeV. In our
case for $\delta m^2 = -1$ eV$^2$ and any small $\sin 2 \theta \ll 1$
we deal with temperatures above $10$ MeV for all essential momenta.
And only for $|\delta m^2| < 10^{-6}$ eV$^2$ one will need to take into
account a more accurate expression for $\rho_{tot}$.}
$\rho_{tot} \simeq 10.75
\pi^2 T^4/30$, the expressions for $U$, $V$, and $Z$ become:
\be
U &=& 1.12\cdot 10^9\cos 2\theta \delta m^2 {x^2\over y}+26.2{y\over x^4},\\
V &=&\frac{29.6}{x^2},\\
Z &=& 10^{10}\left (\eta_{o}
- \int \frac{dy}{4 \pi^2} ~y^2 f_{eq} ~a_-\right )~,
\label{VandZ}
\ee
where $\eta_o$ is the asymmetry of the other particle species (see
eqs. (\ref{etanue},\ref{etanumu})) normalized in the same way as the
neutrino asymmetry (the second term in (\ref{VandZ})). Here we have
implicitly assumed that $\nu_a=\nu_e$.

We can use total leptonic charge conservation to determine $\gamma_-$,
but as we see in what follows, the $\gamma_-$-terms are either sub-dominant
or not important, so we do not need a concrete expression for $\gamma_-$.
The most unpleasant contribution, which makes it so difficult to solve the
symmetric equations numerically, comes from the term containing an
{\it integral} over momentum of the difference $(\rho_{aa}-\bar\rho_{aa})$
with a large coefficient. If the lepton asymmetry is
sufficiently small, $\eta< 10^{-7}$, this term can be
neglected in the symmetric equations, but when it is large its back-reaction
on the rise of the asymmetry  is quite important. All other (asymmetric)$^2$
terms in the symmetric equations, e.g. $\gamma_- a_-$, can always
be neglected. The only essential terms are proportional to $VZ$ because they
enter with a numerically large coefficient (see below).

Let us introduce some more notations. Since the asymmetric term $a_-$ enters
the expression (\ref{VandZ}) with the coefficient $10^{10}$ we introduce
capital letters for the renormalized asymmetric functions:
\be
S=10^{10} s_-,\quad
A=10^{10} a_-,\quad H = 10^{10} h_-,\,\,
{\rm and}\,\,\, L=10^{10} l_-~.
\label{AHL}
\ee
We will also introduce a new variable:
\be
\tau = \xi x^3/y~,
\label{eta}
\ee
where $\xi \approx 6.5 \cdot 10^3 \,\sqrt{|\delta m^2|\cos2\theta} $
so that $U$ vanishes at $\tau =1$ or, in other words,
the MSW resonance takes place at $\tau=1$ if $\delta m^2 <0$ and if the
contribution from the asymmetric part, $VZ$, can be neglected. We will divide
everything by the factor $ M = 1.12 \cdot 10^9 \,  \cos 2\theta \, |\delta m^2| \,x^2/y$, so
that the coefficient functions now become:
\be
F=-\tan 2\theta \approx -\sin 2\theta,\quad U=1/\tau^{2} -1,\quad \gamma =\delta/\tau^2 ~,
\label{newf}
\ee
where $\delta \approx 1/135$ is a small coefficient. In what follows we will
often use the notation $\gamma \equiv \gamma_+$. The asymmetric potential,
$V/M$, in terms of these variables has the form:
\be
V =3.3\cdot 10^{-3} (\cos 2\theta\,\delta m^2)^{-1/3} y q^{-4/3} ~.
\label{newv}
\ee
where we have introduced $q \equiv y\tau$. The equations for the
asymmetric functions can now be written as:
\be
S'/Q &=& F L ~, \label{s'q} \\
A'/Q &=& - F L - 2\gamma A -2\cdot 10^{10} \gamma_- a_+ ~, \label{a'q}\\
H'/Q &=& U L - \gamma H - 10^{10}V Z l_+ - 10^{10} \gamma_- h_+~, \label{h'q}\\
L'/Q &=& \frac{F}{2}(A - S) - U H - \gamma L +
10^{10}VZ h_+ - 10^{10} \gamma_- l_+ \label{l'q} ~,
\ee
where prime now means derivative with respect to $\tau$ and:
\be
Q \approx 5.6 \cdot 10^4\, \sqrt{|\delta m^2 \cos 2\theta}|~.
\label{Q}
\ee
Due to conservation of leptonic charge the integrated contribution of the
last two terms in the r.h.s. of eq.~(\ref{a'q}) vanishes:
\be
\int dy y^2 f_{eq} (y) \left(\gamma A +
10^{10} \gamma_- a_+ \right) = 0~.
\label{consl}
\ee
In the equations for $H'$ and $L'$ we neglect the terms proportional
to $10^{10}\gamma_- \sim Z$  as well as $F(A-S)/2$ because they are small in
comparison with $10^{10}V Z\sim 10^7 Z$.

We will solve these equations in the limit of large $Q\gg 1$, the corrections
generally being of the order of $1/Q$. A formal solution of
equations (\ref{h'q}) and (\ref{l'q}) (in terms of unknown functions $Z$,
$h_+$, and $l_+$) is:
\be
L &=& \frac{10^{10} V Z}{\gamma^2 + U^2} \, \left (
l_+ U + h_+ \gamma \right) ~, \label{Lasymeqn}\\
H  &=&  \frac{10^{10} V Z}{\gamma^2 + U^2} \, \left (
-l_+  \gamma  + h_+ U \right)~.
\label{Hasymeqn}
\ee
This approximation works if $Z(q)$ does not decrease too fast with increasing
$q$ and is even better justified if $Z(q)$ is a rising function of $q$.

These solutions can now be inserted into the set of equations for the
symmetric function, which can be written in the matrix form as
${\cal V}' = {\cal M} {\cal V}$:
\be
\left( \begin{array}{c} s'_+ \\ a'_+\\ h'_+\\ l'_+ \end{array} \right)
= Q\left( \begin{array}{cccc}
0  & 0 & 0 & F\\
0  & -2 \gamma & 0  & -F\\
0 & 0 & -\tilde \gamma &  \tilde U\\
-F/2 & F/2 &  - \tilde U & - \tilde \gamma    \end{array}  \right)
\left( \begin{array}{c} s_+ \\ a_+ \\ h_+ \\ l_+  \end{array} \right)~.
\label{symmatrix}
\ee
Here we have used:
\beq
\tilde U = U \, (1 - D^2/ \sigma^2) \, \, \, \mbox{and} \, \, \,
\tilde \gamma = \gamma \, (1 + D^2/\sigma^2)~,
\eeq
with $D^2 = Z^2 V^2$ and $ \sigma^2 = \gamma^2 + U^2$.

We will solve this system of equations expanding the solution in terms of
the eigenvectors of the corresponding matrix ${\cal M}$ in the r.h.s.
of eq.~(\ref{symmatrix}) with
the coefficients $b_j$ that are not constants (as they would be if
${\cal M}$ was a constant matrix), but functions of $\tau$. For the
functions $b_j$  we
will obtain a set of differential equation that can easily be solved
numerically and even analytically in the limit of a small $\sin 2\theta$.
The matrix ${\cal M}$ has 4 eigenvectors with the eigenvalues:
\be
\mu_1 &\approx & -{F^2 \tilde \gamma \over 2 \tilde \sigma^2} \label{mu1} \\
\mu_2 & \approx & -2\gamma + {F^2(2\gamma -\tilde\gamma) \over 2
[ (2\gamma -\tilde \gamma)^2 + \tilde U^2]} \label{mu2}\\
\mu_{3,4} &\approx & - \tilde \gamma \pm i \tilde U \label{mu34}
\ee
where $\tilde \sigma^2 =\tilde \gamma^2 + \tilde U^2$. The correction
to $\mu_2$ is of the order of $F^2$
when $D^2-\sigma^2 \neq 0$. When the latter quantity is
close to zero, the correction may be of the order of $F$.

The matrix elements of the symmetric density matrix can be expressed through
the four new functions $b_j (\tau)$ as:
\be
s_+ &=& - b_0 +b_1 {F^2\over 4\sigma^2} \nonumber \\
&&- F {\sigma^2 -D^2 \over \sigma^2 \tilde\sigma^2}
\left[\tilde\gamma \left( b_2 \cos \Omega -b_3 \sin \Omega\right) -
\tilde U \left(b_2 \sin \Omega + b_3\cos \Omega\right)\right] ~,
\label{sb}  \\
a_+ &=& - b_0 {F^2(\sigma^2 + D^2) \over 4 \sigma^2 \tilde\sigma^2} +
b_1 {\sigma^2 -D^2 \over\sigma^2} \nonumber \\
&&-{F\over \sigma^2}\left[\gamma \left( b_2 \cos \Omega -b_3 \sin \Omega\right)
+U \left(b_2 \sin \Omega + b_3\cos \Omega\right)\right] ~,
\label{ab}  \\
h_+ &=& b_0 {FU (\sigma^2 -D^2 )\over 2\sigma^2 \tilde\sigma^2} +
b_1 {FU \over 2 \sigma^2} +
{\sigma^2 -D^2 \over\sigma^2} \left(b_2 \sin \Omega + b_3\cos \Omega\right)~,
\label{hb} \\
l_+ &=& b_0 {F\gamma(\sigma^2 + D^2 ) \over 2 \sigma^2 \tilde\sigma^2} -
b_1 {F\gamma \over 2 \sigma^2} +
{\sigma^2 -D^2 \over\sigma^2} \left(b_2 \cos \Omega -b_3 \sin \Omega\right)~,
\label{lb}
\ee
where $\Omega' = Q\tilde U$. To the leading order in the small
parameter $F$ the function $b_0$ satisfies the equation\footnote{Note
that this equation can be written as the evolution equation for the
sterile neutrino (and anti-neutrino) density in momentum space as shown
in ref.~\cite{foot97} (eq. (93)) or ref.~\cite{bell99} (eq. (72)).}
\be
b_0' = - {Q F^2 \tilde\gamma \over 2\tilde\sigma^2} b_0~.
\label{b0'}
\ee
We usually neglect terms of the order of $F^2$, but the one above
contains the large factor $Q$ and is therefore taken into account.

The function $b_1$ is small, $b_1 \sim F^2$, and can be neglected. The
functions $(b_{2,3} \psi )$ satisfy the equations:
\be
\left( b_2 \psi \right)' &=& -Q\tilde\gamma \left( b_2 \psi \right) -
{b_0\over 2} \left[ \left(\tilde\alpha \psi \right)' \sin\Omega +
\left(\tilde\beta \phi \right)' \cos\Omega \right]~,
\label{b2'} \\
\left( b_3 \psi \right)' &=& -Q\tilde\gamma \left( b_3 \psi \right) -
{b_0\over 2} \left[ \left(\tilde\alpha \psi \right)' \cos\Omega -
\left(\tilde\beta \phi \right)' \sin\Omega \right]~,
\label{b3'}
\ee
where $\psi = (\sigma^2 - D^2)/\sigma^2$, $\phi = (\sigma^2 +
D^2)/\sigma^2$, $\tilde \alpha = FU /\tilde \sigma^2$, and $\tilde
\beta= F\gamma /\tilde \sigma^2$.  The initial conditions are $b_0(0)
=-1$ and $b_{1,2,3} (0) =0$.

\begin{figure}[t]
\begin{center}
\psfig{file=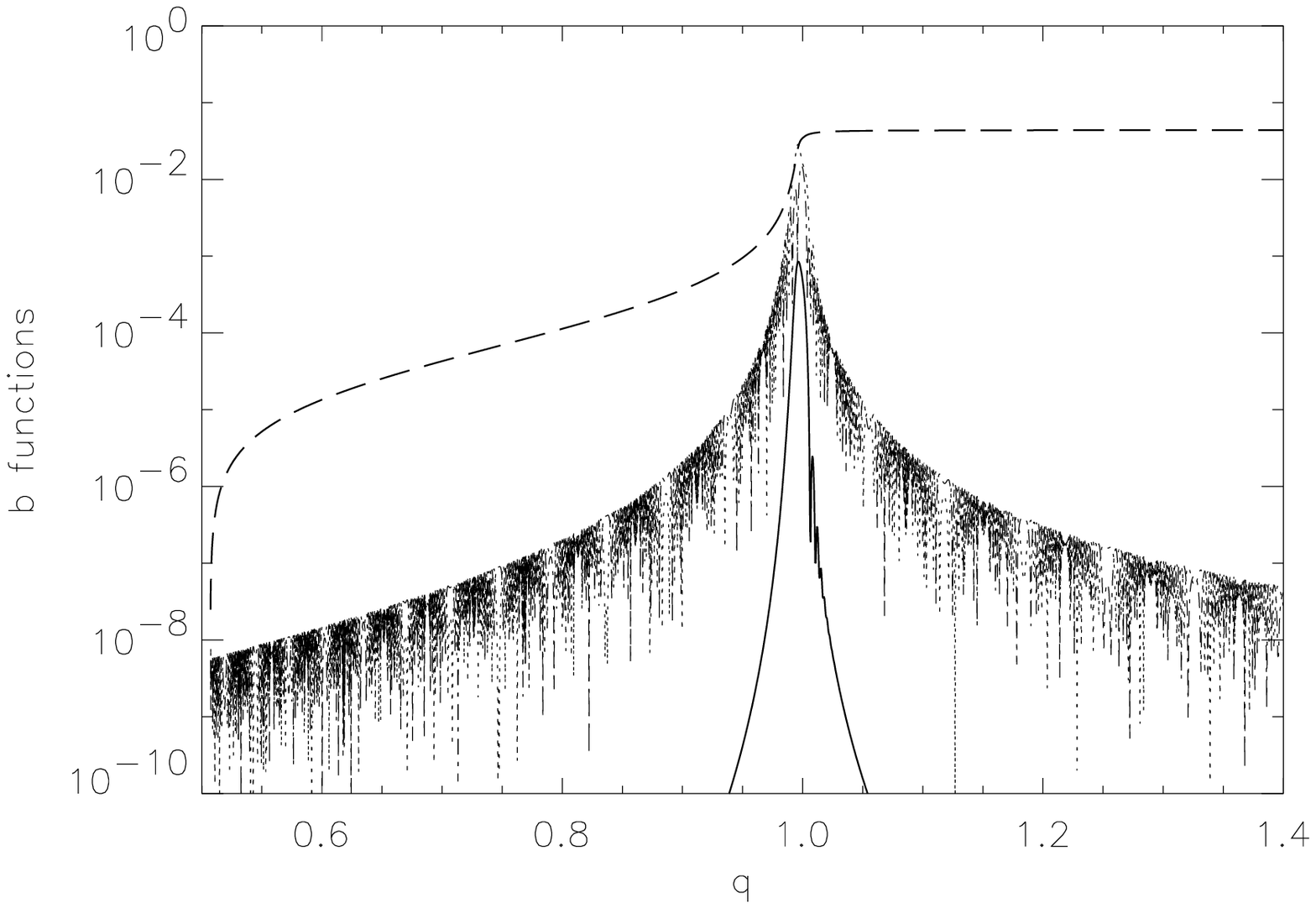,width=5in,height=3.2in}
\end{center}
\caption{}
\label{osc:fig1}
\end{figure}

When $D=0$ it is straightforward to solve for the $b$-functions
numerically. From fig.~\ref{osc:fig1} it is clear that $b_{1,2,3}$ are
very small except near the resonance. For momentum $y=1$ and
parameters sin$2\theta=10^{-3}$ and $\delta m^2 = -1$, we see that the
function $b_0$ follows the curve $\exp \left[-\kappa\cdot
(\mbox{arctan}(2(q-1)/\delta) +\pi/2 )\right]$ to a high accuracy
(remember that $q = y\tau$).  Here $\kappa$ goes like sin$^22\theta$
for small mixing angles.

The last two equations~(\ref{b2'},\ref{b3'}) can be solved as:
\be
b_2 \,\psi &=& -{1\over 2} \int_0^\tau d\tau_1 e^{-\Gamma_3(\tau) +
\Gamma_3(\tau_1)} b_0 (\tau_1) \left[ \left( \tilde\alpha \psi \right)'_1
\sin \Omega_1 + \left( \tilde \beta \phi \right)'_1 \cos \Omega_1 \right]~,
\label{b2psi} \\
b_3 \,\psi &=& -{1\over 2} \int_0^\tau d\tau_1 e^{-\Gamma_3(\tau) +
\Gamma_3(\tau_1)} b_0 (\tau_1) \left[\left(\tilde\alpha \psi \right)'_1
\cos \Omega_1 - \left( \tilde \beta \phi \right)'_1 \sin \Omega_1 \right]~,
\label{b3psi}
\ee
where $\Gamma_3' = Q \tilde \gamma$ and sub-1 means that the argument of
the corresponding function is $\tau_1$. When substituting these results into
eqs.~(\ref{hb},\ref{lb}) and integrating by parts  we obtain:
\be
h_+ &=& {Q\over 2} \int_0^\tau  d\tau_1 e^{-\Delta\Gamma
 } b_0 (\tau_1) \left[ F\sin \Delta\Omega  -
{b_0' \over Q b_0} \left( \tilde\beta_1 \phi_1 \sin  \Delta\Omega  -
 \tilde\alpha_1 \psi_1 \cos \Delta\Omega  \right) \right]~,
\label{lfin} \\
l_+ &=& -{Q\over 2} \int_0^\tau  d\tau_1 e^{-\Delta\Gamma
 } b_0 (\tau_1) \left[ F\cos \Delta\Omega  -
{b_0' \over Q b_0} \left( \tilde\alpha_1 \psi_1 \sin \Delta\Omega  +
 \tilde\beta_1 \phi_1 \cos \Delta\Omega \right) \right]~,
\label{hfin}
\ee
where $\Delta \Gamma_3 =  \Gamma_3 (\tau) - \Gamma_3 (\tau_1)$ and
$\Delta\Omega = \Omega(\tau_1) - \Omega(\tau)$. The integrals can be taken
in the limit of large $Q$ according to:
\be
\int_0^\tau d\tau_1 \Phi (\tau_1) e^{-\Delta\Gamma -i \Delta\Omega} =
{1\over Q}{\Phi (\tau) \over \tilde \gamma -i\tilde U}~.
\label{intphi}
\ee
This result permits us to express the function $L$ from
eq.~(\ref{Lasymeqn}) algebraically through the
lepton asymmetry $Z$:
\be
L = - 10^{10} { F V Z \gamma U \over \sigma^2 \tilde\sigma^2} b_0\left[
1- {F^2(\sigma^2 + D^2) \over 4\sigma^2 \tilde\sigma^2} \right]~.
\label{Lfin}
\ee
Substituting this result into eq.~(\ref{a'q}) and integrating it with
$dy y^2 f_{eq}(y)/4\pi^2$ we finally derive the following equation
governing the evolution of the lepton asymmetry $Z(q)$:
\be
\frac{1}{Z}\frac{d Z}{dq} = - \delta\, B   q^{5/3} \int_0^\infty dt \,
{ t^4 (t^2-1) f_{eq} (tq) b_0(1/t)
\over \sigma^2 \tilde\sigma^2 } \left( 1-B_1 { \sigma^2 +D^2 \over
\sigma^2 \tilde\sigma^2 } \right)~,
\label{dzdq}
\ee
where:
\be
B &=& 4.71\cdot 10^4 (\cos 2\theta |\delta m^2|)^{1/6} \left({\sin 2\theta \over
10^{-3}} \right)^2~,
\label{B} \\
B_1 &=& 2.5 \cdot 10^{-7}\left({\sin 2\theta \over 10^{-3}} \right)^2~,
\label{B1}
\ee
and where we have introduced the new integration variable $t =1/\tau$,
which is proportional to the neutrino momentum. If we neglect the term
proportional to $B_1$ our evolution equation (\ref{dzdq}) coincides
with the main contribution of the ``static approximation'' in
refs.~\cite{foot96,foot97} (see for instance eq.~(65) in
\cite{bell99}). However the last term in eq.~(\ref{dzdq}) stops the
rise of the asymmetry $Z$ earlier and at much lower values.

In the limit of a small $Z$ we may neglect $D$ in the r.h.s. of this
equation and it can easily be integrated. If we assume that $b_0$ does
not vary, i.e.  $b_0 \equiv -1$, then the integral over $dt$ is
(approximately) proportional to $(1-q)$, so that for $q<1$ the
asymmetry decreases, and for $q>1$ it starts to rise. It is easy to
check that with $D=0$ the integrated rise is stronger than the
decrease and thus the asymmetry rises by the factor $\exp\, [3.7
(10^5\sin 2\theta )^2]$ for the mass $\delta m^2 = -1$. For any $\sin
2\theta \geq 2\cdot 10^{-5}$ there would be an enormous rise of the
asymmetry.

This does not happen, however, because for a large $\theta$ the variation
of $b_0$ should be taken into account. It changes as:
\be
\Delta b_0 = b_0 (\infty)- b_0 (0)
\approx 0.04 \left(\frac{\sin 2\theta}{10^{-3}}\right )^2 |\delta m^2|^{1/2}~,
\label{Deltab0}
\ee
Even this relatively weak
variation happens to be vitally important for the evolution of the asymmetry.
The point is that the integral in the r.h.s. of eq.~(\ref{dzdq}) has a
resonance
at $U=0$. The contribution of this resonance into the integral is quite small
for a constant $b_0$, because the factor $U$ in the numerator cancels it out
since $U$ is an odd function near the resonance.
However, since $b_0$ experiences variation exactly at the resonance point, its
variation breaks the symmetry from the positive and negative contribution
of $U$ near the
resonance and the relative effect of the small variation of $b_0$
is enhanced by the factor $1/\delta\sim 10^2$. This effect diminishes the
positive contribution into the r.h.s. of eq.~(\ref{dzdq}) and  the
rise of the asymmetry is strongly suppressed. In particular for
$\sin 2\theta \geq 10^{-3}$ the integrated contribution of the r.h.s. becomes
negative and the asymmetry decreases with respect to its initial value.

The variation of $b_0$ is also important because it gives an upper limit
to a possible generation of lepton asymmetry. Because of leptonic charge
conservation the asymmetry generated in the sector of active neutrinos
must be equal to that in the sector of sterile ones. The latter is
proportional to the difference of the diagonal matrix elements of the density
matrix, $\Delta Z \sim \Delta (s-\bar s)$. Since the variation of $s$ and
$\bar s$ can only be positive (initial value of both is $-1$),
$\Delta (s-\bar s)<\Delta (s+\bar s)$ and the last quantity is given by
the variation of $b_0$. Naively taken from eq.~(\ref{Deltab0}) this variation
is rather small for small $\theta$. However, as we see in what follows,
the variation of $b_0$ is rising with rising $Z$, so the discussed limit
is not broken. On the other hand, the back-reaction from the variation of
$b_0$ terminates the rise of the asymmetry when it is still not too large; the
maximum amplification, that happens to be near
$\sin 2\theta \approx 10^{-5}$ for $\delta m^2 = -1$, could be about $10^6$.

For $10^{-5}<\sin 2\theta < 10^{-3}$ a solution of eq.~(\ref{dzdq})
without back-reaction results in a huge rise of the asymmetry,
however, the back-reaction efficiently kills this rise and the
asymmetry may increase at most by $6$ orders of magnitude. It
confirms the early assertion (based on oversimplified arguments) of
ref.~\cite{barbieri91} that back-reaction does not permit the
asymmetry to grow too much. However, the magnitude of the generated
lepton asymmetry the we found is much larger than that advocated in
ref.~\cite{barbieri91} but still much smaller than in
refs.~\cite{foot96}--\cite{shi96}, except for a very large mass
difference, $\delta m^2 \approx 10^9$ eV, where the asymmetry may be
above $0.1$. We have solved eq.~(\ref{dzdq}) numerically in the above
mentioned range of $\sin 2\theta$.  For sufficiently small values of
$q$ the equation was solved directly without any simplifications,
while for larger $q$, when the product $\zeta=3.3\cdot 10^{-3} Z (\cos
2\theta \delta m^2)^{-1/3}$ became close to unity, the integral was
estimated in the resonance approximation. There are two resonances
corresponding to the condition $D^2 =U^2$ that give opposite sign
contributions to the integral, and eq.~(\ref{dzdq}) becomes:
\be
\zeta' = \sum_{j=1,2} {\pi B\over 2} b_0\left( 1/t_j\right)
{q^{8/3} (t_j^2 -1) f_{eq} (qt_j) \over \zeta \sqrt{ 4q^{2/3} +\zeta^2}}
\left[ 1 - B_1 {(t_j^2 -1)^2\over \delta^2 t_j^4 } \right]~,
\label{zeta'}
\ee
where $t_j = \pm \zeta /2q^{1/3} + \sqrt{ 1 + (\zeta /2q^{1/3})^2}$.

\begin{figure}[b]
\begin{center}
\psfig{file=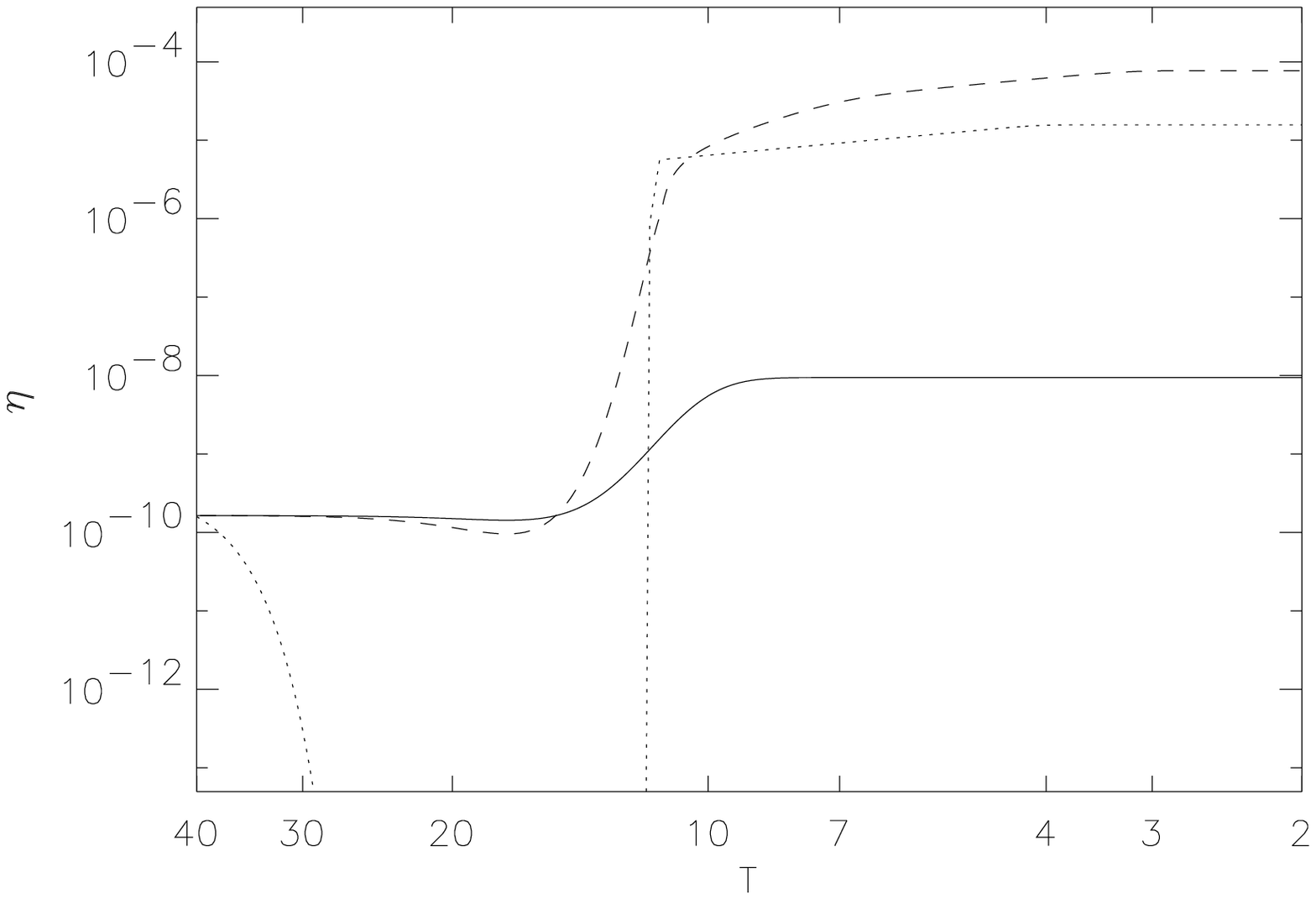,width=5in,height=3.2in}
\end{center}
\caption{}
\label{osc:fig2}
\end{figure}

For $\delta m^2 = -1$ we present the evolution of $\eta$ as a function
of the decreasing temperature $T$ for 3 different mixing angles in
fig.~\ref{osc:fig2}, where the lepton asymmetry is
$\eta=|\eta_B/4-(n_{\nu_e}-n_{\bar{\nu}_e})/n_\gamma|$. The solid line
is for $\sin 2\theta = 1 \cdot 10^{-5}$, and it is clearly seen that
the asymmetry is frozen at a very low value.  For bigger mixing
angles, sin$2 \theta = 2 \cdot 10^{-5}$ (dashed) or sin$2 \theta = 3
\cdot 10^{-4}$ (dotted) the increase may be much bigger.  The dotted
line clearly shows a power law behavior, $\eta
\propto T^{-1}$, following the exponential increase. If one neglects the
back-reaction, $B_1=0$, then the power law becomes $\eta \propto
T^{-11/3}$.

\begin{figure}[t]
\begin{center}
\psfig{file=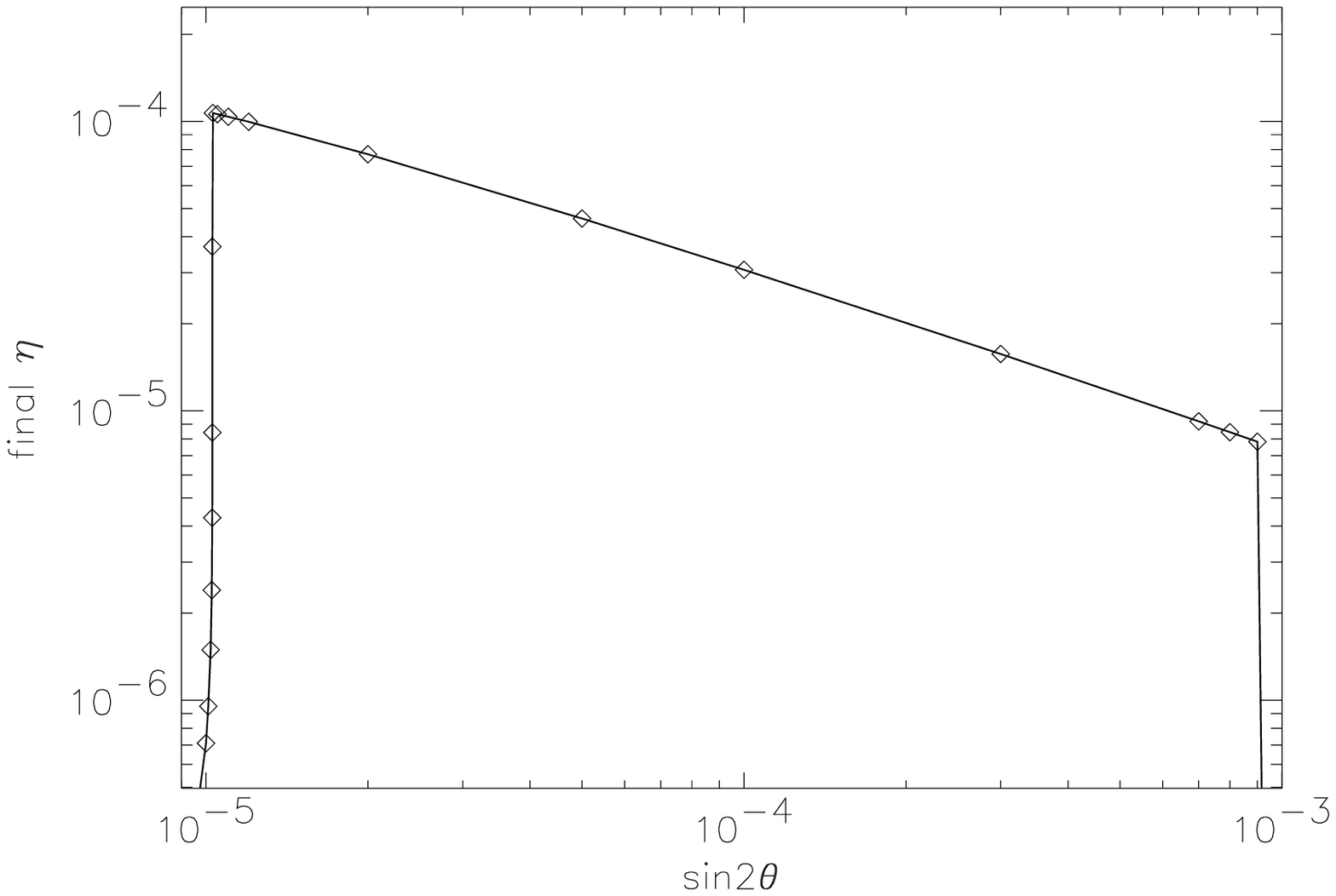,width=5in,height=3.2in}
\end{center}
\caption{}
\label{osc:fig3}
\end{figure}

In fig.~\ref{osc:fig3} we plot the final value of $\eta$ as a function
of sin$2\theta$ for $\delta m^2 = -1$. One clearly sees the sharp
exponential cut-off around sin$2\theta=10^{-3}$. The final lepton
asymmetry in the region with a large increase, $10^{-5} <
\mbox{sin}2\theta <10^{-3}$, does not depend on the initial value,
$\eta_{in}$, whereas the final value of $\eta$ is almost linear in
$\eta_{in}$ for $\mbox{sin}2\theta <10^{-5}$.

\begin{figure}[b]
\begin{center}
\psfig{file=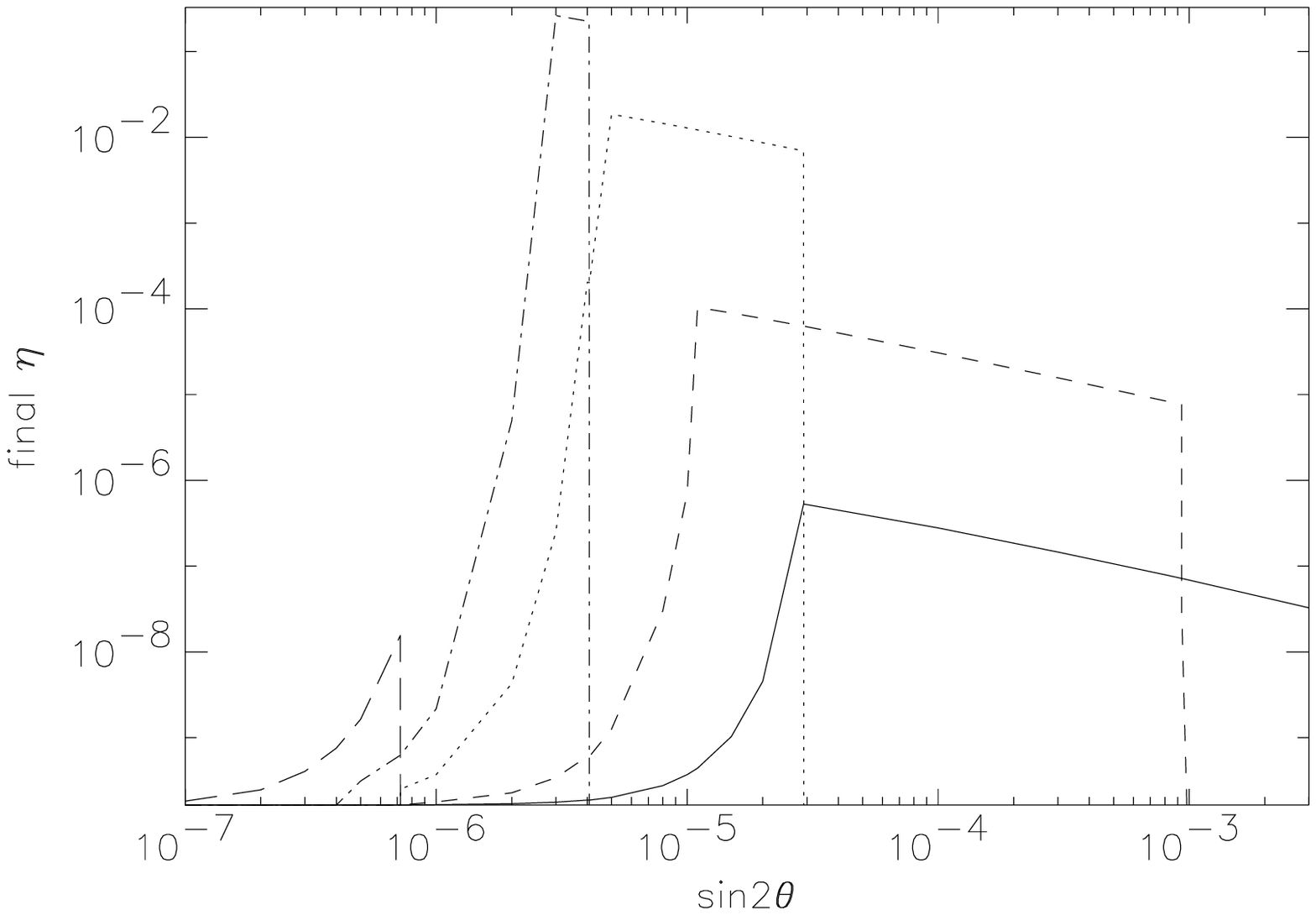,width=5in,height=3.2in}
\end{center}
\caption{}
\label{osc:fig4}
\end{figure}

\begin{figure}[t]
\begin{center}
\psfig{file=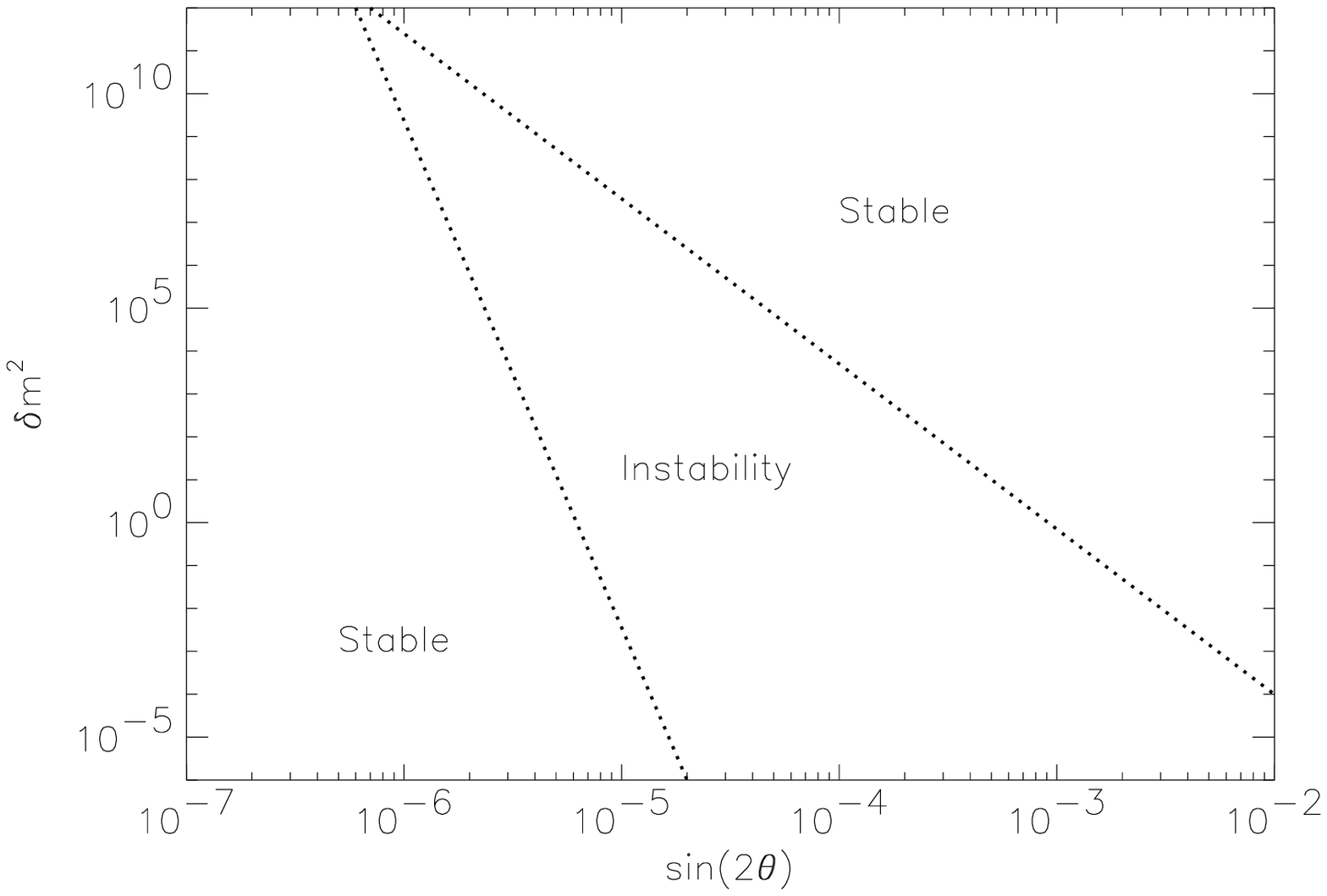,width=5in,height=3.2in}
\end{center}
\caption{}
\label{osc:fig5}
\end{figure}

For smaller masses the region of increase is shifted slightly to
higher mixing angles as is seen in fig.~\ref{osc:fig4}, where we plot
the final lepton asymmetry, $\eta$, as functions of sin$2\theta$ for 5
different masses, $-\delta m^2 = 10^{-6}, 1, 10^6, 10^{9}, 10^{12}$.
This shift is caused by $B \sim \left( |\delta m^2| \right)^{1/6}$
(see eq.~(\ref{B})).  On the other hand, for bigger masses the
exponential cut-off moves to smaller mixing angles. This is because
$\Delta b_0$ goes like $\sqrt{|\delta m^2|}$.  Clearly the effect with
the biggest masses has limited applicability, since very heavy
particles would become non-relativistic early, however, it is
comforting to note that the maximal asymmetry generated does not
continue rising for very heavy neutrinos. The region of instability in
the ($\delta m^2,\sin 2\theta$)-plane is presented in
fig.~\ref{osc:fig5}.

Our eq.~(\ref{dzdq}) is very similar to the equations describing the
evolution of the asymmetry derived in ref.~\cite{bell99}. However, we
took terms related to the variation of $b_0$ into account and these
terms are responsible for the stabilization of the rise of the
asymmetry when the latter is still small.  Our derivation of the
evolution equation is somewhat different and to our mind it is more
rigorous.

The result are only valid in the case of small mixing angles, $\sin
2\theta < \delta = 1/135$. In the other limiting case the evolutionary
equation is quite different as well as the behavior of the
asymmetry. Our preliminary results show that in the case of a large
mixing angle the asymmetry does not rise.  This agrees with the
result found in this work that for $\sin 2\theta >10^{-3}$ the
asymmetry diminishes. Even though the asymmetry remains small, the
impact of neutrino oscillations on the primordial nucleosynthesis
could be non-negligible due to the effects of non-equilibrium neutrino
distribution function. We will calculate the abundances of light
elements in a subsequent paper.

Thus, according to our calculations, lepton asymmetry in the sector of
active neutrinos may indeed be strongly enhanced. However the
enhancement that we found is considerably weaker than that found in
the earlier papers \cite{foot96}-\cite{footAP99}.  Even for a very
large mass difference $-\delta m^2 = 10^6$ the resulting
asymmetry is below $10^{-2}$ and only for $-\delta m^2 = 10^9$
it reaches the values that may be
important for nucleosynthesis (see fig.~\ref{osc:fig4}). We
believe that our calculations are very accurate.  The only
approximation is an expansion in inverse powers of $Q$
 (eq.~(\ref{Q})) and in powers of a small $\sin 2\theta <0.01$.
Otherwise our calculations are exact. In some
cases we used analytic solutions to appropriate differential equations
but in all the cases numerical solutions were quite simple and they
agree very well with the analytic results. In the limit of a small
$\delta m^2$, e.g.  $-\delta m^2 = 10^{-7}$ the parameter $Q\approx 18$
still remains large.  Our results for such small $\delta m^2$ are in a
reasonable agreement with direct numerical calculations made in
refs.~\cite{kirilova97} and with our own ones (unpublished).

We see that for some values of the mixing angle the asymmetry $Z$
first very quickly goes to zero reaching extremely small values about
or below $10^{-100}$ and later started to rise up to $10^{-5}$ or even
somewhat larger. If one solves eq.~(\ref{dzdq}) the sign of the final
asymmetry remains fixed and is completely determined by the initial
conditions. So in this approximation we do not observe any chaoticity
in agreement with earlier papers~\cite{foot97,dibari99}.  On the other
hand if one applies a direct numerical approach, then it is
practically evident that chaoticity must be observed because no direct
computational procedure is able to maintain an accuracy at the level
$10^{-100}$. Thus one would expect that in the region when the
asymmetry is quite small its numerically calculated sign is arbitrary
and chaotic; it is just numerical errors. When the asymmetry starts to
rise its final sign is the same as the initial one at the moment
when the asymmetry becomes larger than the numerical accuracy.  It
could possibly explain the chaoticity observed in the
papers~\cite{shi96,shi99,enqvist99}.

However one should remember that eq.~(\ref{dzdq}) is valid only if $Z$
does not decrease too fast with increasing $q$. So for a small $Z$ one
cannot say on the basis of eq.~(\ref{dzdq}) that the asymmetry is as
small as $10^{-100}$. There are some more terms in
eqs.~(\ref{s'q}-\ref{l'q}), as e.g. $F(A-S)/2$, that should not be
neglected if $Z$ vanishes. Our preliminary results show that the
solution of the kinetic equations in the limit of small $Z$ does not
show any chaoticity, though the sign of $Z$ may be different from the
initial one. We do not observe the sign change but this statement
demands some further checks.

There is a physically interesting possibility of chaoticity, namely
if the asymmetry, as calculated through kinetic equations, is
extremely small, the statistical fluctuations would be essential. The
relative magnitude of a statistical fluctuation in a volume with $N$
particles is about $1/\sqrt N$. So if this value is larger than the
asymmetry $Z$ the fluctuations would dominate and the sign of the
asymmetry would be determined by statistical fluctuations. However, to
be essential the size of the region with such a fluctuation should be
larger than the neutrino diffusion length during the characteristic
time of oscillations. The complexity of the calculations in such a case
would increase very much because now one has to take into account the
effects of the fluctuating medium on the oscillations.

A possible explanation of the difference between our results and the
results of other groups (they also disagree between themselves, in
particular in a possible chaotic behavior of the asymmetry)
is that in most cases an assumption of kinetic equilibrium
of neutrinos was made. This assumption enormously simplifies the
numerical calculations but may be strongly violated. Its violation may
be crucial for the strength of generation of lepton asymmetry. We
checked in a simplified example that in the opposite limit when the
spectrum of neutrinos never recovered its equilibrium distribution and
the resonance is complete (i.e.  for relatively large $\sin 2\theta$),
the asymmetry experiences only a very mild enhancement~\cite{dolgov99}.

However in some papers (see e.g. \cite{footAP99}) it is stated that a
complete set of kinetic equations was numerically solved without any
approximations. It is always difficult to find the source of
disagreement, especially in numerical works. As we understood from the
paper~\cite{footAP99}, the system of $8N$ kinetic equations (where $N$
is the number of points in the momentum grid) for $\nu$ and $\bar\nu$
(equivalent to our eqs.~(\ref{firstasyms})--(\ref{firstasyml})) was solved
numerically but an additional equation for the evolution of lepton
asymmetry was introduced ($Z$ in our notations and $L$ in notations of
the quoted paper). The latter was obtained from the expression for the
asymmetry by differentiating the corresponding integrand containing
elements of the density matrix. This equation was also solved
numerically step by step and the resulting asymmetry was substituted
into the equations describing the evolution of the density matrix
elements. Possibly this technical trick helped to diminish the
computational instability of the original equations. To calculate the
integral the authors estimated it close to the resonance and
integrated over the range of 3.5 resonance widths. We repeated a similar
procedure for the derivative of the asymmetry $Z'$ in the resonance
approximation and found both analytically and numerically that the
result strongly depends upon the integration limits. The larger are
the limits the weaker is the rise of asymmetry. The integral becomes
saturated when the limits are larger than 10 or even 20 widths. Such a
slow rate of saturation is related to the fact that the derivative of
the resonance is a function that changes sign near the resonance.
Another source of disagreement may be that the integration over
momentum performed in ref.~\cite{footAP99} was symmetric around the
resonance, however in reality the range of integration in negative and
positive directions are not the same (because the momentum runs from $0$
to $\infty$). The term of the lowest order in the resonance width in
the integrand is an odd function near the resonance so the
contribution of asymmetry in the integration range is enhanced by
$1/\delta$. This effect is not strong in the case when the lepton
asymmetry remains small and its back-reaction is not essential but
when it starts to rise, the asymmetry in integration limits should be
taken into account. These two effects may possibly explain the
difference between our results and the calculations of ref.~\cite{footAP99}.
However in our discussion with R. Foot and R. Volkas they defended stability
of their calculations with respect to the choice of the region of integration.
So at the moment the question about the precise origin of our
disagreement remains open.

In conclusion, we have analytically transformed the complete set of
momentum dependent equations governing the evolution of the neutrino
distribution functions to a form which allows a simple numerical
solution. The only approximation is an expansion in the small
parameter sin$2 \theta$. These equations can even be solved
analytically in the limit of large $Q$, allowing us to derive a simple
first order differential equation for the evolution of the lepton
asymmetry.  This differential equation takes into account the strong
back-reaction effects on the generation of the lepton asymmetry due
to the presence of an extra term (proportional to $B_1$) which is
absent in approximate equations derived in some other papers. Due to
this back-reaction we find that the asymmetry rise terminates 
at a much smaller magnitude.

\bigskip

{\bf Acknowledgments}

The work of AD and SH was supported in part by the Danish National
Science Research Council through grant 11-9640-1 and in part by
Danmarks Grundforskningsfond through its support of the Theoretical
Astrophysical Center. SP was supported by SISSA and by the TMR network
grant ERBFMRX-CT96-0090. DS was supported in part by the Russian
Foundation for Fundamental Research through grant 98-02-17493-A. DS
and SP thank TAC for hospitality when part of this work was done.  We
acknowledge our discussion with P.~Di Bari and R.R.~Volkas during
COSMO-99 where this work was presented. 
We exchanged several e-mail letter with P. Di Bari, R. Foot, and R. Volkas
and though we did not reach an agreement, the discussion is very much
appreciated.

%
%
\nc{\advp}[3]{{\it  Adv.\ in\ Phys.\ }{{\bf #1} {(#2)} {#3}}}
\nc{\annp}[3]{{\it  Ann.\ Phys.\ (N.Y.)\ }{{\bf #1} {(#2)} {#3}}}
\nc{\apl}[3] {{\it  Appl. Phys. Lett. }{{\bf #1} {(#2)} {#3}}}
\nc{\apj}[3] {{\it  Ap.\ J.\ }{{\bf #1} {(#2)} {#3}}}
\nc{\apjl}[3]{{\it  Ap.\ J.\ Lett.\ }{{\bf #1} {(#2)} {#3}}}
\nc{\app}[3] {{\it  Astropart.\ Phys.\ }{{\bf #1} {(#2)} {#3}}}
\nc{\cmp}[3] {{\it  Comm.\ Math.\ Phys.\ }{{ \bf #1} {(#2)} {#3}}}
\nc{\cqg}[3] {{\it  Class.\ Quant.\ Grav.\ }{{\bf #1} {(#2)} {#3}}}
\nc{\epl}[3] {{\it  Europhys.\ Lett.\ }{{\bf #1} {(#2)} {#3}}}
\nc{\ijmp}[3]{{\it  Int.\ J.\ Mod.\ Phys.\ }{{\bf #1} {(#2)} {#3}}}
\nc{\ijtp}[3]{{\it  Int.\ J.\ Theor.\ Phys.\ }{{\bf #1} {(#2)} {#3}}}
\nc{\jmp}[3] {{\it  J.\ Math.\ Phys.\ }{{ \bf #1} {(#2)} {#3}}}
\nc{\jpa}[3] {{\it  J.\ Phys.\ A\ }{{\bf #1} {(#2)} {#3}}}
\nc{\jpc}[3] {{\it  J.\ Phys.\ C\ }{{\bf #1} {(#2)} {#3}}}
\nc{\jap}[3] {{\it  J.\ Appl.\ Phys.\ }{{\bf #1} {(#2)} {#3}}}
\nc{\jpsj}[3]{{\it  J.\ Phys.\ Soc.\ Japan\ }{{\bf #1} {(#2)} {#3}}}
\nc{\lmp}[3] {{\it  Lett.\ Math.\ Phys.\ }{{\bf #1} {(#2)} {#3}}}
\nc{\mpl}[3] {{\it  Mod.\ Phys.\ Lett.\ }{{\bf #1} {(#2)} {#3}}}
\nc{\ncim}[3]{{\it  Nuov.\ Cim.\ }{{\bf #1} {(#2)} {#3}}}
\nc{\np}[3]  {{\it  Nucl.\ Phys.\ }{{\bf #1} {(#2)} {#3}}}
\nc{\pr}[3]  {{\it  Phys.\ Rev.\ }{{\bf #1} {(#2)} {#3}}}
\nc{\pra}[3] {{\it  Phys.\ Rev.\ A\ }{{\bf #1} {(#2)} {#3}}}
\nc{\prb}[3] {{\it  Phys.\ Rev.\ B\ }{{{\bf #1} {(#2)} {#3}}}}
\nc{\prc}[3] {{\it  Phys.\ Rev.\ C\ }{{\bf #1} {(#2)} {#3}}}
\nc{\prd}[3] {{\it  Phys.\ Rev.\ D\ }{{\bf #1} {(#2)} {#3}}}
\nc{\prl}[3] {{\it  Phys.\ Rev.\ Lett.\ }{{\bf #1} {(#2)} {#3}}}
\nc{\pl}[3]  {{\it  Phys.\ Lett.\ }{{\bf #1} {(#2)} {#3}}}
\nc{\prep}[3]{{\it  Phys.\ Rep.\ }{{\bf #1} {(#2)} {#3}}}
\nc{\prsl}[3]{{\it  Proc.\ R.\ Soc.\ London\ }{{\bf #1} {(#2)} {#3}}}
\nc{\ptp}[3] {{\it  Prog.\ Theor.\ Phys.\ }{{\bf #1} {(#2)} {#3}}}
\nc{\ptps}[3]{{\it  Prog\ Theor.\ Phys.\ suppl.\ }{{\bf #1} {(#2)} {#3}}}
\nc{\physa}[3]{{\it Physica\ A\ }{{\bf #1} {(#2)} {#3}}}
\nc{\physb}[3]{{\it Physica\ B\ }{{\bf #1} {(#2)} {#3}}}
\nc{\phys}[3]{{\it  Physica\ }{{\bf #1} {(#2)} {#3}}}
\nc{\rmp}[3] {{\it  Rev.\ Mod.\ Phys.\ }{{\bf #1} {(#2)} {#3}}}
\nc{\rpp}[3] {{\it  Rep.\ Prog.\ Phys.\ }{{\bf #1} {(#2)} {#3}}}
\nc{\sjnp}[3]{{\it  Sov.\ J.\ Nucl.\ Phys.\ }{{\bf #1} {(#2)} {#3}}}
\nc{\sjp}[3] {{\it  Sov.\ J.\ Phys.\ }{{\bf #1} {(#2)} {#3}}}
\nc{\spjetp}[3]{{\it Sov.\ Phys.\ JETP\ }{{\bf #1} {(#2)} {#3}}}
\nc{\yf}[3]  {{\it  Yad.\ Fiz.\ }{{\bf #1} {(#2)} {#3}}}
\nc{\zetp}[3]{{\it  Zh.\ Eksp.\ Teor.\ Fiz.\ }{{\bf #1} {(#2)} {#3}}}
\nc{\zp}[3]  {{\it  Z.\ Phys.\ }{{\bf #1} {(#2)} {#3}}}
\nc{\ibid}[3]{{\sl  ibid.\ }{{\bf #1} {#2} {#3}}}
%
%

\newpage
{\large \bf Figure Captions:}
\vskip0.5cm
\noindent
{\bf Fig. 1} $~~~$ $b_j$ as functions of $q$ for momentum $y=1$, sin$2
\theta = 10^{-3}$ and $\delta m^2 = -1$.  The long-dashed line is
$1+b_0$, the full line is $b_1$, the dotted and the dashed cures are
absolute values of $b_2$ and $b_3$ respectively.

\noindent
{\bf Fig. 2} $~~~$ The evolution of $\eta$ as a function of the
decreasing temperature $T$ in MeV. The full line is for sin$2 \theta =
1 \cdot 10^{-5}$, the dashed line is for sin$2 \theta = 2 \cdot
10^{-5}$, and the dotted line is for sin$2
\theta = 3 \cdot 10^{-4}$. All with $\delta m^2 = -1$.

\noindent
{\bf Fig. 3} $~~~$ The final value of $\eta$ as a function of sin$2
\theta$ for $\delta m^2 = -1$. For mixing angles bigger than $\approx
10^{-3}$ the final value of $\eta$ is exponentially suppressed (see
text). We have used $\delta m^2 = -1$, $\eta_{in} \sim 10^{-10}$ and 
$q$ runs from $10^{-2}$ to $10^3$.

\noindent
{\bf Fig. 4} $~~~$ The final value of $\eta$ as a function of
sin$2\theta$ for 5 different masses: $-\delta m^2 = 10^{-6}$ (solid),
$1$ (dashed), $10^6$ (dotted), $10^9$ (dash-dot), and $10^{12}$
(long-dashed).

\noindent
{\bf Fig. 5} $~~~$ Stability regions in $\sin 2\theta-\delta
m^2$ space. Here {\it instability} means that the final $\eta$ is more
than an order of magnitude bigger than the initial $\eta$.

\newpage

\end{document}